# Rethinking Information Theory for Mobile Ad Hoc Networks[1]

Jeff Andrews, Nihar Jindal, Martin Haenggi, Randy Berry, Syed Jafar, Dongning Guo,
Sanjay Shakkottai, Robert Heath, Michael Neely, Steven Weber, Aylin Yener

**Abstract**

The subject of this paper is the long-standing open problem of developing a general *capacity theory* for wireless networks, particularly a theory capable of describing the fundamental performance limits of mobile ad hoc networks (MANETs). A MANET is a peer-to-peer network with no pre-existing infrastructure. MANETs are the most general wireless networks, with single-hop, relay, interference, mesh, and star networks comprising special cases. The lack of a MANET capacity theory has stunted the development and commercialization of many types of wireless networks, including emergency, military, sensor, and community mesh networks. Information theory, which has been vital for links and centralized networks, has not been successfully applied to decentralized wireless networks. Even if this was accomplished, for such a theory to truly characterize the limits of deployed MANETs it must overcome three key roadblocks. First, most current capacity results rely on the allowance of unbounded delay and reliability. Second, spatial and timescale decompositions have not yet been developed for optimally modeling the spatial and temporal dynamics of wireless networks. Third, a useful *network* capacity theory must integrate rather than ignore the important role of overhead messaging and feedback. This paper describes some of the shifts in thinking that may be needed to overcome these roadblocks and develop a more general theory that we refer to as *non-equilibrium information theory*.

## 1 The Need for a Network Information Theory

Information theory has provided a scientific foundation for the development of some of society's most advanced and beloved technologies including computers, cellular phones, and the Internet. It seems natural that technologies such as these will converge, allowing ubiquitous wireless connectivity. However, the performance limits of the decentralized wireless networks of the future are presently not known, and optimal approaches to designing these networks are known

---

only for a few special cases. The most challenging and general class of wireless networks to both quantify and design are *ad hoc networks*, which are mobile, peer-to-peer networks that operate without the assistance of pre-existing infrastructure. Immediate applications include emergency and battlefield networks, metropolitan mesh networks for broadband Internet access, and sensor networks. In addition to these pending applications, this important open problem is quite general and a solution to it will likely impact the science of networks in other fields, including biology, economics, and air and automobile transportation.

A central concept in information theory is *capacity*, which is the boundary between the physically possible and physically impossible in terms of reliable data rate. For a given transmitter and receiver, the link capacity for Gaussian noise channels is well-defined and given by the well-known formula $C = B\log_2(1+SNR)$, where $B$ is the bandwidth, and *SNR* is the signal to noise ratio. This simple formula – often known as the *Shannon limit* – and its implications have been indispensable in the development of today's vast communications infrastructure. By providing a target, it has encouraged large investment in developing high speed communications. Perhaps more importantly, the insights provided by information theory have often provided a roadmap to communication engineers. Communication link rates now approach the Shannon limit even in challenging time-varying wireless channels. In time-varying channels capacity becomes a random variable, and so statistical properties of the Shannon limit like ergodic (average) and outage capacity are typically used.

This success has not yet translated to wireless *networks*, which, for $K$ mobile devices, comprise $K(K-1)$ possible one-way connections – without including multicasting. In fact, the general Shannon limit is not known even for $K = 3$ with static channels, due to difficulties in modeling the interactions between the 6 possible one-way links [CovElG79, Kra05]. In ad hoc networks, $K$ can be on the order of 10, 100, or even 1000, and all the links are time-varying. Classical link-based information theory does not appear well-suited to the role of describing network performance limits, any more than understanding the functionality of a single neuron gives insight on how the brain at large functions or a single transistor's behavior characterizes the behavior and capabilities of a modern CPU.

## 2   The Three Roadblocks

Despite the obvious difficulties in adapting a fundamentally point-to-point theory to a network, there have been numerous efforts at extending information theory to networks, often referred to as multi-terminal or network information theory. Such extensions have proven to be extremely difficult for most cases of interest, which has motivated considerable work on capacity scaling laws that attempt to describe how the end-to-end achievable rates in the network scale as a function of the total number of nodes in it. An overview of the progress in these directions can be found in [XueKum06]. This paper will argue that even if straightforward extensions of classical information theory to $K > 2$ were successful, there are three fundamental roadblocks to a *functional* network information theory that are not addressed by the Shannon framework. By *functional*, we mean a network information theory that provides useful upper bounds on end-to-end network throughput. These upper bounds should be robust to nonidealities in the capacity model, encompass a notion of timescales and delay, and point to limits that may be approached in the foreseeable future with arbitrarily good engineering.

A network information theory that addresses these three roadblocks and provides useful capacity limits for MANETs is likely to require significant innovation relative to the contemporary Shannon framework.

**Roadblock 1:  Network capacity requires different foundational assumptions than link-based information theory.**

The link capacity expression $C = B\log_2(1+SNR)$ was revealed by considering memoryless channels with arbitrarily long blocklengths (delay) and vanishingly small error probability. Surprisingly, allowing unbounded delay, reliability, and complexity did not wind up compromising the usefulness of this result in links – indeed, very high reliability can be achieved at rates near the Shannon limit, and due to design and processing advances over the past several decades, capacity-approaching strategies now have a delay and complexity that is acceptable for many applications. However, MANETs have bursty traffic sources, end-to-end delay constraints that are much more difficult to meet, and mobility which is constantly changing the network topology. In a link, delay is primarily related to the codeword length, and delays on the order of thousands of channel symbols are, from a capacity point of view, close enough to infinity for the asymptotic limits to be accurate while still providing a link that can be close to "real-time" in terms of human perception. In networks, delays are measured on much larger timescales corresponding to buffer times, traffic patterns, channel access times, multihop routing, retransmissions, and user mobility. The delays required for asymptotic limits to be meaningful in the context of maximizing network throughput might be on the order of tens of seconds, minutes, or even longer, which is orders of magnitude larger than permissible delay bounds. As a consequence, the stochastic variations in the channels, queues, and routes due to fading, mobility, and traffic patterns cannot be averaged out and have to be explicitly considered. Table 1 summarizes the approximate timescales required for different algorithms in MANETs. In nearly all cases, slower time-scale dynamics allow more sophisticated techniques to be used at all layers of the network stack.

In summary, the capacity of a MANET is closely related to the timescales occurring in the network, which are driven by external dynamics. What is needed is a *non-equilibrium* information theory, which rather than averaging over all the dynamics, is capable of describing capacity in the context of local equilibria.

**Roadblock 2: Wireless networks defy familiar link-based decompositions. New decompositions need to account for nodal interactions over time and space.**

Information theory has been tremendously successful in analyzing centralized wireless systems because such networks can generally be decomposed into constituent links. Traditional Shannon theory applies to these links, and the system as a whole can be characterized as the aggregation of these links. For example, cellular systems can be first decomposed into individual cells, which comprise a point-to-multipoint (broadcast/downlink) and multipoint-to-point (multi-access/uplink) channel. These multiuser channels can be further decomposed via orthogonalization (in time, frequency, space, or code) into point-to-point links. In most cases, although less often recently, interference from other links can be simply treated as additional noise. The modeling of cellular systems is relatively robust to this idealization due to the careful layout of cell sites and the single-hop communication paradigm. A further functional decomposition can be performed that allows network "layers" to be treated separately: the physical layer is thought of as a bit pipe while the higher layers exist to provide the physical layer with bits to transmit. In practice, these layers have been designed and optimized separately, with a few recent exceptions such as opportunistic scheduling and network coding.

MANETs evade such decompositions due to their decentralized structure, the need for multihop routing, the dynamic traffic and network topologies, and the inherent coupling between channels with multiple transmitters and receivers interleaved in space. Therefore, the familiar layered network stack – which is a decomposition of system functionality – also needs to change, possibly adaptively. Ideally, each layer should be defined by the timescale over which it operates, so "higher" layers (whose state changes slowly) can reasonably interact with equilibrium states of the layers below it, whose states change more quickly.

Although this decomposition roadblock is fairly well recognized, MANET research and design has still generally followed the traditional separation of network and physical layer functionalities. A majority of MANET research from the network community is performed on the basis of overly simplified physical layer models; nodes can communicate at a prescribed rate if they are within "range", and cannot otherwise. In contrast, information theory has essentially ignored networking concerns like bursty traffic, finite flows and sessions, queuing delay, and routing. Though convenient and familiar, neither of these approaches sufficiently captures the level of interaction between these functionalities that occur in MANETs. Indeed, network coding, one of the field's more revolutionary recent ideas, came about precisely by departing from the traditional "packet as atomic unit" perspective [AhlCai00].

**Roadblock 3: Overhead is a much greater burden in MANETs and must be accounted for in the capacity theory itself.**

A considerable amount of overhead is left unaccounted for by information-theoretic channel and network models [Gal76]. Even general point-to-point communication models start off by making many implicit assumptions: a connection has been established (often using a separate control channel), synchronization has been achieved, packet headers relating to addressing and other overhead information have been sent. In a link, such simplifications may be viable, because these costs are either relatively minor, easily accounted for in a lump sum manner (e.g. "network overhead" of 20%), or may be non-recurring and hence amortized over the lifetime of the link. However, in a network, accomplishing these tasks may consume extensive system resources, and defy a simple characterization. This is not simply an academic problem; current military prototype MANETs routinely experience overhead on the order of even 99% of the end-to-end packet transmissions [Mar04, Swa06].

In a dynamic network, the cost of maintaining optimal communication routes may be severe at every layer of the network stack. In addition to the synchronization, channel estimation, route selection, handshaking and standard acknowledgement messaging required in an MANET, many promising schemes – relaying, cooperative diversity, beamforming, opportunistic scheduling, and backpressure routing – require substantial real-time overhead. If a MANET has capacity $C$ when each node has perfect knowledge of all state information in the network, then the best effective capacity that can possibly be achieved is $C^* = C' - R_{OH}$, where $C' < C$ is the capacity of the network with partial state information, and $R_{OH}$ is the cost of achieving that partial state information[2]. It is important to understand when state information acquisition improves the $C^*$; that is, when is overhead messaging justified by a net capacity increase? This requires a careful accounting of the overhead implied by design decisions. Developing a unifying analytical framework to account for overhead messaging is critical for a relevant network information theory.

## 3 Functional Capacity: Capacity with Constraints

Capacity is primarily a mathematical concept. Capacity, when unconstrained, allows for arbitrary delay, reliability arbitrarily close to 1, and unlimited computational complexity. The reason capacity has had operational relevance in the design of communication systems is because the delay and complexity needed to approach it have turned out to be reasonable in current technology. Similarly, although Shannon promises "perfect" reliability, error probabilities of 1 in a million or less are typical of current systems, which passes for "perfect" when coupled with upper layer quality assurance measures such as CRC checks and ARQ retransmissions. There has been research on intermediate delay and reliability regimes, for example reliability functions, error exponents, finite-length coding, and general rate-reliability limits via information spectrum [VerHan94].

---

[2] We do not mean to imply that capacity is necessarily a scalar. Denoting it as such here and later in the paper is for conceptual purposes.

Information theory, in full generality, is certainly capable of handling non-asymptotic regimes for delay, reliability, and for that matter any constraints. What has allowed information theory to produce tractable and revealing capacity expressions is *ergodicity*, which allows disturbances to be averaged out of over time and the equilibrium behavior of the channel (or network) to be determined. Because of the considerable dynamics in networks, using ergodicity as the foundation for a MANET information theory is dubious, and key tools of information theory such as the law of large numbers, the asymptotic equipartition property, and Stein's Lemma may not be suitable in many scenarios of interest.

One problem with developing a suitable information theory for MANETs has been the difficulty in balancing required constraints with the well-justified view that excessive restrictions might preclude a true upper bound on performance[3]. Consider the left half of Figure 1: Shannon capacity (unconstrained) is the upper bound, the ultimate boundary between the physically possible and physically impossible. The *functional capacity* corresponds to what might be achievable with great engineering under a tenable set of assumptions. The *functional capacity* is a realistic upper bound on system performance, which allows for large but not unbounded delay, low but not vanishingly small probability of error, very high but not unlimited signal processing complexity, sufficient but not unlimited feedback. The functional capacity of course lies below the Shannon limit. In links – even in wireless links – these curves wind up being very close to each other. The *achievability* curve, which might be achieved by a state-of-the-art system, is below both of them, but again by a small amount. This allows the functional capacity to be ignored in links, and so capacity can be thought about simply in terms of just *optimality* (the Shannon limit) and *achievability* (a system design that approaches the Shannon limit).

---

[3] For example, the *cutoff rate* was widely believed to be the achievable limit in AWGN channels; a belief disproven by turbo codes and their descendants, which very closely approach the Shannon limit.

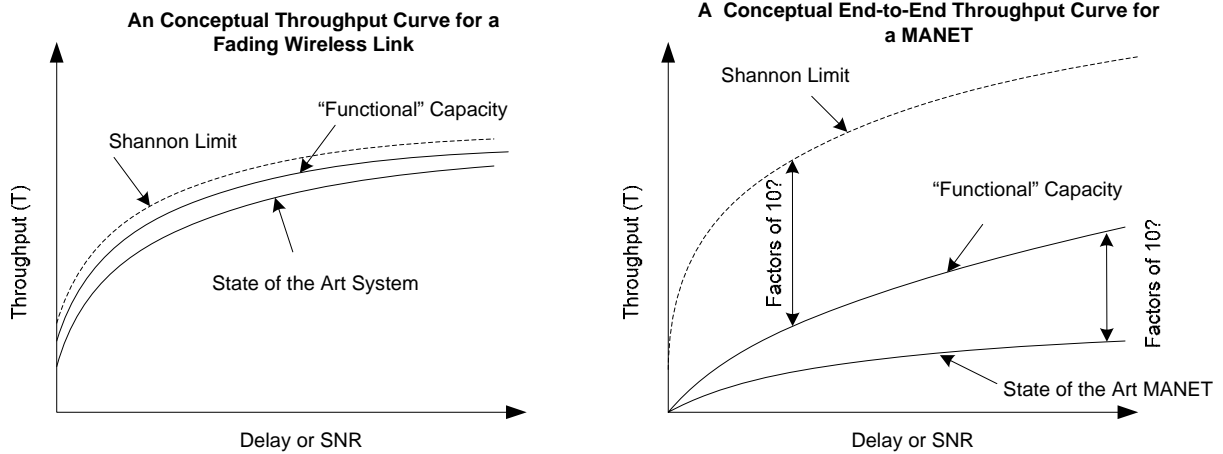

Figure 1: A conceptual example of what Throughput vs. Delay or SNR might be for a link (left) and for a MANET (right).

In MANETs however, the situation is very different. Although both the Shannon limit and functional capacity of MANETs are unknown, we believe the functional capacity will not even be close to the Shannon limit in most cases. In contrast to links, the Shannon limit does not provide a meaningful upper bound for MANETs. Even if the Shannon limit was known, it would not tell us much about the ultimate performance potential of MANETs. We suspect that current state-of-the-art MANET designs are also far below the functional capacity, although by how much we are not sure. The reasons for these large gaps, in contrast to the link case, center around the network dynamics, the complicated geography of the interference, the intense overhead demands that mobility places on all levels of the network stack, and the lack of viable centralized control for scheduling and routing. The key observation is that ultimate Shannon limits on the performance will likely be extremely optimistic in networks as opposed to links, even with arbitrarily good engineering many years into the future. Therefore, if placing (reasonable) constraints on the Shannon limit makes it easier to compute, this might actually be a good thing.

We concede that the notion of functional capacity is bothersome, and reeks of defeatism and subjectivity. It opens the door to all manner of debate as to what would comprise a "functional" theory. In many respects, it goes square against the beauty of information theory, which establishes a speed of light to shoot for, while leaving the engineering to the engineers.

Nevertheless, the characteristics of MANETs are such that some movement in this direction is unavoidable if we want to describe heterogeneous network performance limits in a meaningful way, and be able to answer questions that are of interest to engineers. For example, given a certain bandwidth, network topology, mobility and traffic time-scales, and delay constraint, what would be the maximum set of throughput pairs that one can ever hope to achieve in this network?

In short, functional capacity can be thought of as an important special case of information theory, where judiciously applied constraints provide a theory that is robust to nonidealities, provides useful insights, and with luck, is more tractable. There are numerous precedents for exploring capacity with additional constraints, for example, constraints on peak and average power, amount of channel state feedback, and delay. For illustrative purposes we now give two examples – one from signal processing and one from networking – that show why such a step is particularly critical for MANETs.

**Example 1: Interference cancellation and dynamic range.** A widely adopted constraint is that a radio cannot simultaneously transmit and receive in a single frequency band because the ratio of the transmit to receive powers ($P_t/P_r$) is enormous, in the range of 50-100 dB. Mathematically though, a known transmit signal can be subtracted from the received signal, permitting simultaneous transmission and reception. In practice this is impossible, since even the slightest cancellation error appears amplified by $P_t/P_r$. Several classes of multiuser channels have been shown to achieve capacity with some form of interference cancellation. Although alternative achievability strategies may exist, a MANET capacity result that depends heavily on interference cancellation should be viewed with some level of skepticism, since the dynamic ranges of different *received* signals are much larger than in centralized networks due to the irregular geometry of the users. Can a receiver perfectly cancel out an interferer that is 50 dB stronger than its desired transmitter? Information theory says "yes". Reality – due to finite bandwidth A/D converters, the impossibility of perfect channel estimation, and the lack of infinite precision computing – says "no". To what extent should information theory bend to accommodate implementation realities? Are some implementation realities fundamental while others may change with time?

**Example 2: Mobility and infinite delay**. As a thought experiment consider an ad hoc network with *K* nodes, where each node has some mobility pattern that ensures that over an infinite time horizon it will travel arbitrarily close to every location in the network. It can therefore be assured that every node will at some point be close enough to every other node for a reliable transmission to take place, regardless of the interference from simultaneous transmissions elsewhere. With buffers at each node, this random connectivity can be exploited by routing data to available nearby nodes, and then having each intermediate node act as a "postman" by delivering the data when the desired destination is encountered. This allows the end-to-end throughput to be limited only by the rate at which each node encounters other nodes, rather than by the rate at which each node encounters its destination [GroTse02]. However, the postman model of routing incurs a very large delay, and hence is impractical in applications with even modest delay constraints. This example – which has great capacity but only if users can wait a few hours or days for their data – illustrates the danger of neglecting delay when discussing the capacity of networks.

The above two examples show how the Shannon framework – without appropriate additional constraints – can result in upper bounds that are extremely sensitive to nonidealities. In contrast, the Shannon framework is extremely robust in links, and nonidealities such as non-Gaussian noise and codebooks do not typically change the main insights gained from the theory. The central argument for *functional capacity* is that in its current form, Shannon's framework does not provide a robust platform for MANET capacity. Like the Shannon limit for links, a functional MANET information theory will provide a target to what can actually be achieved with great engineering, and should avoid degenerate cases like the above two examples where the capacity may appear large, but is in fact very fragile.

## 4   Non-Equilibrium Information Theory

In classical information theory, a communication channel is modeled via one or more random processes modeling quantities such as the channel gain and additive noise. A key reason Shannon was able to develop an elegant theory is that by allowing for arbitrarily long delays one can appeal to various laws of large numbers and thus average over these processes. Essentially, with long enough codes, each code word experiences the steady-state or *equilibrium* behavior of the channel. Furthermore, in practice, the delays needed to reasonably approach this equilibrium behavior have proven to be acceptable.

**Throughput, Delay, and Reliability (TDR) Regions**

In classical (equilibrium) information theory, the capacity of a multi-terminal system is characterized by the maximum reliable throughput between terminals, called the capacity region. A useful non-equilibrium information theory would likely characterize a network by other metrics in addition to throughput. In particular, two other fundamental quantities are *delay* and *reliability*. These are both needed here because by definition arbitrarily long delays are not allowed, and without arbitrarily long delays, achieving arbitrarily high reliability is not possible. Furthermore, these three metrics are the key performance measures to most users and applications. We refer to them as a MANET's TDR (throughput-delay-reliability)-triplet. These quantities are inter-related, i.e. a session will typically be able to achieve higher reliability by reducing its throughput or increasing its delay. The performance metric of interest is the set of achievable TDR values for each session in a network. Of course, coupling will also exist between the TDR values for multiple sessions. An interesting question is to which extent it is possible to characterize the TDR region for a source-destination pair in a network given the TDR regions for all individual links, or small component networks.

We have been somewhat cavalier in our use of throughput, delay and reliability so far. There are several possibilities for exactly how these quantities are defined. For example, one possibility is that the throughput T is the total amount of data received at the intended destination by delay D with probability R, meaning that on average RT bits are reliably received per unit time. Delay could also be measured in terms of the maximum or average delay per bit, while reliability could be measured at the bit level or packet level. Note, here we are talking about end-to-end quantities in a MANET. There have been a number of efforts at calculating or bounding such quantities in wire-line networks, which may be useful in developing such a theory for MANETs. Examples include (stochastic) network calculus and large deviations [ShwWei95].

In a non-equilibrium setting, the achievable TDR values will be time-varying due to slower timescale dynamics. Since uncertainties cannot be averaged out over the duration of a typical session or the lifetime of a route, MANETs are almost permanently in a transient state, pursuing evasive equilibria. Since the transient phases are dominant, classical information theory is less relevant, and new analysis tools are needed that explicitly account for the dynamics. In other words, *infostatics* (i.e., classical information theory) is not sufficient for the characterization of dynamic systems such as MANETs – they require *infodynamics*. Analogously, the emergence of thermodynamic and electrodynamic systems necessitated the development of the thermodynamic and electrodynamic theories based on their static counterparts. Infodynamics is a synonym for a non-equilibrium information theory, i.e., a theory the explicitly characterizes non-asymptotic regimes (in terms of time, reliability, and network size).

## 5 The Way Forward

It may appear that developing new foundations for a non-equilibrium information theory is well beyond reach. However, we find hope in several directions. First we note that there have already been limited applications of such non-equilibrium ideas in information theory. For example, the notion of outage capacity can be viewed as a non-equilibrium theory for a fading channel, where here reliability corresponds to the outage probability. A key idea in this work is to essentially assume a separation of timescales so that certain randomness (i.e. the additive noise) is averaged over while other randomness (i.e. the fading) is not. Such an approach will likely be even more useful in developing a non-equilibrium theory for MANETs, where the performance at a given timescale can be treated as the expected performance conditioned on the realization of all dynamics at slower timescales. Understandably the difference between some of the timescales may not be enough to warrant the use of laws of large numbers, in which case innovative techniques need to be developed to provide a succinct interface for interactions between timescales.

We now briefly overview some recent research directions that we think hold potential for better understanding the capacity of wireless networks.

**Lessons from Physics**. Wireless networks are fundamentally physical systems, governed by the laws of physics. Rather than assuming a particular mathematical channel model, Franceschetti [Fra07] recently combined information theory with electromagnetic propagation laws and found that efforts to beat Gupta and Kumar's scaling law are fruitless since Maxwell's equation fundamentally limit the degrees of freedom available in the network. If a communication system with many degrees of freedom (in time, space and/or frequency) is modeled as a thermodynamic system, the Shannon capacity is a statistical phase transition point, beyond which arbitrarily low error probability is impossible. Statistical physics methodologies, such as the replica method, have been successfully applied to obtain the capacity of multiuser and MIMO systems. Furthermore, statistical physics offers a number of modeling tools for dealing with non-equilibrium systems and large quantities of random variables. Additionally, non-equilibrium statistical mechanics, which studies macroscopic systems in which the dominance of statistics disappears, provides a rich collection of relevant theory and experience. Moreover, the theories that physicists have developed for dynamic interacting many-particle systems may become relevant [Lig04] as well as the models and tools used for the analysis of vehicular traffic [Chow00]. In particular, the microscopic theories of vehicular traffic are promising since they do not treat vehicle flows as compressible fluids but explicitly focus on the dynamics of the individual vehicle. In other words, the theories treat the vehicular network as a system of interacting particles driven far from equilibrium, enabling the study of the dynamics of more general non-equilibrium systems – such as MANETs.

**Random Graphs and Stochastic Geometry.** Another promising, underutilized toolset is the rich theory of random graphs, stochastic geometry, and percolation theory. An inherent feature of ad hoc wireless networks is that users are randomly located, as are source-destination pairs and possible relay nodes. Since the path loss is the dominant effect of both desired and interference power in a wireless network, the positions of the nodes is inseparable from the capacity of the network. If the nodes are located in an i.i.d. manner either through a random scattering or because of mobility, their spatial distribution is well modeled by a Poisson point process. Rich toolsets on these subjects have been developed and are under development by mathematicians [Sto96,Boll06], including for non-Poisson point processes. These tools allow interference distributions and outage probabilities to be explicitly derived in closed-form, which allows connectivity and spatial throughput to be quantified precisely. An exact analysis of these quantities is possible in the special case when the node locations are Poisson, channels fading is Rayleigh, and medium access control is uncoordinated [Bac06]. Achieving good approximations when one or more of these assumptions are relaxed is the subject of ongoing work [WebAnd07, GanHae07].

**Capacity Approximation Techniques**. Since exact capacity characterizations may be impossible for MANETs, capacity approximations may be the key to understanding the performance limits of wireless networks. Promising recent ventures in this direction include the degrees of freedom approach and the deterministic channel approach. The degrees of freedom of a network provides a capacity approximation that is accurate within $o(\log(SNR))$ and in some cases within $O(1)$ [CadJaf07]. Deterministic channel models have led to capacity approximations accurate to within a few bits in several interesting cases [AveDig07]. These approximations share a common philosophy: since interference rather than thermal noise will be the principal bottleneck to wireless network performance, it is useful to de-emphasize noise and focus on the interactions of the desired signals and interference terms.

The idea of interference alignment that emerged out of the degrees of freedom perspective has revealed the fallacy of the "cake-cutting" interpretation of orthogonal spectrum allocation. An interesting example of interference alignment shows that it is possible for everyone to use half of the channel resource with no interference to one another. The key to this counter-intuitive result is the realization that the alignment of dimensions is relative to the observer (the receiver), and since each receiver has a different perspective it is possible to simultaneously satisfy seemingly contradictory spectrum requirements. Interference alignment schemes constructed on the deterministic channel illuminate the close relationship between it and the degrees of freedom perspective approach. Interference alignment also reaffirms the observation that structured codes are needed for network capacity theorems [NazGas07]. While it is known that both structured (lattice) and random codes can achieve capacity on the point to point channel, information theorists have mostly relied on random codes to come up with achievability schemes for capacity theorems. For networks it seems this may not be the right approach. Intuitively, in a network, a code designed for one user designs the interference to another user. Since random codes will not automatically align themselves, structured codes may be necessary for wireless networks. Thus, degrees of freedom analyses, deterministic channel models, and achievable schemes based on structured codes in general and interference alignment in particular are promising approaches for estimating the capacity of wireless networks.

**Control Theory**. While robustness is often not treated in information theory, control theory has a rich tradition in doing so. Because optimal control algorithms are often intolerant to changes in the environment or plant, control theorists have developed *robust* control theory. This important branch of control theory deals with changing system parameters and the design of algorithms that exhibit graceful degradation in the presence of changes. In doing so, robust control systems optimize the design space for approaches that can maintain their stability and performance in the face of unpredictable dynamics. In short, "robust control refers to the control of unknown plants with unknown dynamics subject to unknown disturbances" [Cha96]. Similarly, we contend that MANET analysis techniques should not be overly sensitive to changes in the system assumptions. In other words, the *functional* capacity of a MANET should change gracefully with changes in the (distributions of the) relevant modeling parameters. As control theorists have realized, this is a necessary condition for a theory with practical relevance, since real networks can never be modeled exactly. There has been some work in this direction in information theory, perhaps best characterized by considering the capacity of channels when the channel distributions are uncertain [LapNar98]. In general however, the consideration of robustness to assumptions and modeling is not a prominent aspect of contemporary information theory.

# 6 Conclusions

The development of an accurate and robust capacity theory for wireless ad hoc networks is one of the most difficult and important challenges remaining in information theory, and has major ramifications on the fields of wireless networking and communications. Meeting this challenge will require new ideas, new tools, and a willingness to think outside the confines of conventional information theory. In this paper, we have not presented solutions. We have attempted, however, to get closer to asking the right questions. It appears clear that (limited) delay and reliability must be addressed at a fundamental level by any successful networking information theory. Furthermore, understanding the unique spatial and temporal dynamics of ad hoc networks – and accounting for the overhead messaging that they require – will be essential. With hesitation, we wonder what constraints may be needed in order to develop a useful set of upper bounds that may prove attainable in the coming decades. We overviewed promising recent developments in information theory, and suggest possible connections with historically unrelated fields. The development of a non-equilibrium information theory that characterizes – rather than averages over – the effects of dynamics would be one of the most important breakthroughs in communications since Shannon's theory.

# 7  Acknowledgements

The authors would like to thank DARPA and the National Science Foundation for their support of our research. In particular, we would like to thank Chris Ramming of DARPA for his encouragement to write this paper and for his and Richard Barron's ongoing feedback, which has helped us clarify these ideas.

We would also like to thank several colleagues who have provided feedback on these ideas over the past year and half, including Shlomo Shamai, Bob Gallager, Mike Honig, Eytan Modiano, and Andrea Goldsmith and the rest of the "FLoWS" team.

**Table 1: Implications of Different Timescale Coherence Times on MANET capacity**

| *Time-scale Coherence* | *Communication and network algorithms* | *Effects* |
|---|---|---|
| $10^{-6}$ seconds (unlikely) | Non-coherent communication<br>Mobility increases short-term capacity | MANET communication close to impossible |
| $10^{-3}$ seconds (very high mobility) | Coherent communication | $C \approx E[\log(1 + SNR)]$ with coding & interleaving |
| | ALOHA or round robin scheduling | Although fast fading helps multiuser capacity, it is difficult for the transmitters to learn the channels |
| | Flooding | Inefficiency due to very short shelf-life of routes |
| $10^{-2}$–$10^{-1}$ secs (moderate mobility) | Adaptive Modulation and Coding<br>Spatial Diversity<br>Local power control and scheduling | Increase link capacities, exploit channel variations |
| | Handoff algorithms | Can maintain routing, continuity of service |
| | ARQ | Improved link robustness |
| 1 second (low mobility) | Limited Channel-State Feedback, Queue-state feedback | Richer transmitter optimization becomes possible, such as beamforming |
| | Backpressure Routing based on differential queue lengths | Approaches throughput and delay optimality by avoiding congestion |
| | Network Coding | Mix packets for robustness and throughput |
| 10 – 100 secs (no mobility) | Rich channel state feedback: near-optimal MIMO transmission, waterfilling, interference channels known | Capacity on links can be approached, e.g. multiuser MIMO precoding, interference cancellation |
| | Rich network feedback: network-level scheduling and routing | Cost of network-level feedback can be amortized; sophisticated coordination viable |